\documentclass[twocolumn,showpacs,amsmath,amstex,amssymb,mathfonts,prb]{revtex4-1}
\usepackage{graphicx,bm,units}

\begin{document}

\title{Three dimensional phase diagram of disordered HgTe/CdTe Quantum spin-Hall wells}

\author{Emil Prodan}
\address{Department of Physics, Yeshiva University, New York, NY 10016} 

\begin{abstract} We compute the phase diagram of the HgTe/CdTe quantum wells in the 3 dimensional (3D) parameter space of Dirac mass, Fermi level and disorder strength. The phase diagram reveals the Quantum spin-Hall, the metallic and the normal insulating phases. The phase boundary of the Quantum spin-Hall state is shown to be strongly deformed by the disorder. Taking specific cuts into this 3D phase diagram, we recover the so called topological Anderson insulator (TAI) phase, but now we can demonstrate explicitly that TAI is not a distinct phase and instead it is part of the Quantum spin-Hall phase. The calculations are performed with $S_z$-conserving and $S_z$-nonconserving Hamiltonians.
\end{abstract}


\date{\today}

\maketitle

The HgTe/CdTe quantum wells were the first candidates as Quantum spin-Hall (QSH) insulators. Theory predicted that HgTe/CdTe quantum wells can be tuned to enter the QSH phase, where one should observe robust conducting helical channels flowing around the edges of the samples.\cite{Bernevig:2006hl} This prediction was partially confirmed by experiments,\cite{Koenig:2007ko} which showed quantized direct conductance for small contact distances. The conductance measurements, however, consistently showed a decay of the direct conductance with the distance between the contacts, a hallmark of localization. This behavior spurred several studies that looked into the robustness of the edge modes of HgTe/CdTe quantum wells against de-phasing,\cite{Xing:2008lj,JiangPRL2009xi} and strong disorder (references given below).  

While pursuing such program, one study detected a seemingly new phase, which was dubbed the topological Anderson insulating (TAI) phase.\cite{LiPRL2009xi} Like the QSH phase, TAI was found to display perfectly conducting channels at the edges of the samples, in spite of the presence of strong disorder. The parameter region where this phase was observed seemed to be totally disconnected from the other phases, so one could legitimately conclude that TAI is indeed a distinct phase of the system. Specifically, TAI was observed at large disorder, for Dirac masses that would give a topologically trivial system if the disorder were turned off, and for Fermi levels well inside the conduction band of the clean system. These findings were subsequently confirmed by independent studies,\cite{GrothPRL2009xi,JiangPRB2009sf,Yamakage2010xr} thought sometime different interpretations were given.

To be more concrete, let us introduce the relevant model. As shown in Ref.~\onlinecite{Bernevig:2006hl}, the ``low energy" band theory of the clean HgTe/CdTe wells can be captured by a 2D effective Hamiltonian (written in the momentum space):
\begin{equation}
\begin{array}{c}
H_0({\bm k})=\left ( 
\begin{array}{cc}
h({\bm k}) & \Gamma({\bm k}) \\
\Gamma({\bm k})^\dagger & h^*(-{\bm k})
\end{array}
\right ),
\end{array}
\end{equation}
where $h({\bm k})$$=$$\epsilon({\bm k})$$+$${\bm d}({\bm k})$$\cdot$${\bm \sigma}$, with ${\bm \sigma}$=$(\sigma_x,\sigma_y,\sigma_z)$ encoding the Pauli's matrices, and $\Gamma({\bm k})$ is a $S_z$-nonconserving interaction. The behavior of the measured energy bands in the proximity of the $\Gamma$-point can be captured by the following expression for ${\bm d}({\bm k})$:
\begin{equation}
{\bm d}= (A \sin k_x,A \sin k_y, \Delta-2B(2-\cos k_x - \cos k_y)).
\end{equation}
In the absence of disorder and with $\Gamma({\bm k})$ set to zero, $H_0$ displays a topological phase if $0$$<$$\Delta/B$$<$$4$ and $4$$<$$\Delta/B$$<$$8$ (the insulating gap closes when $M/B$=0, 4 and 8), and a topologically trivial phase if $\Delta/B$$<$0 or $\Delta/B$$>$8.

Working within the Born approximation, a second study \cite{GrothPRL2009xi} on disordered HgTe/CdTe quantum wells showed that disorder renormalizes the Dirac mass term $\Delta$ and the chemical potential of the system. The renormalization occurs in the ``right" direction, so as to give rise to the TAI phase. Based on this analysis, this reference concluded that TAI occurs because of the real gap renormalization. 

Numerically, the phase diagram of strongly disordered QSH insulators was computed in the $(E_F,W)$ plane ($W$ = disorder strength) using the transfer matrix approach,\cite{Onoda:2007xo} by probing the conductance of the edges,\cite{LiPRL2009xi,GrothPRL2009xi} and by level statistics analysis and computation of the quantized bulk invariants.\cite{ProdanJPhysA2010xk} The phase diagram was also computed in the $(\Delta,W)$ plane using the transfer matrix method.\cite{Yamakage2010xr} When computed in the $(E_F,W)$ plane, the TAI phase appears to be disconnected (distinct) from the other phases, but when computed in the $(\Delta,W)$ plane, the TAI phase appears to be connected to, or be a part of the QSH phase.

The studies mentioned above provided diverging views about the TAI phase, which led to a splitting of the community following or working on this effect, with one side claiming that TAI is a novel phase and the other side saying that TAI is connected to the QSH phase, and therefore is not a new phase. The issue can be settled down by a computation of the global phase diagram of the system in the 3D parameter space of $(E_F,\Delta,W)$, in which case one can follow the evolution of the phase diagram in all three directions and therefore record the emergence (or the lack of it) of various disconnected phase components. This is exactly what we are set to do in this work.

We have recently developed an efficient numerical method to evaluate the spin-Chern number in the presence of disorder.\cite{Prodan:2009oh,Prodan:2010cz,ProdanJPhysA2010xk,Shulman2010cy} The methodology stems from a combination of the non-commutative theory of the spin-Chern number (applicable to $S_z$-nonconserving models too),\cite{ProdanJPhysA2010xk}  and a novel numerical algorithm to evaluate topological invariants in the real-space representation.\cite{Prodan2010ew,ProdanJPhysA2010xk} In the present work, we will use such computations to map the regions where the spin-Chern number displays quantized values and ultimately to construct the  3D phase diagram of HgTe/CdTe QSH wells in the presence of strong disorder. We have shown in the past,\cite{Prodan:2010cz,ProdanJPhysA2010xk,Shulman2010cy} and we have verified this again, that such calculations are in perfect agreement with other methods of investigations, such as level statistics analysis. 

Based on our results, we can now show explicitly that the TAI phase is the same as the Quantum spin-Hall phase, and that the apparent distinct character of the TAI phase seen in the original paper\cite{LiPRL2009xi} is due to the fact that the phase diagram was examined only along a 2D section of this 3D phase diagram. We will show exactly where that section occurs in our 3D phase diagram. In fact, we will show that our 3D phase diagrams are in good quantitative agreement with all the previous numerical calculations. 

Besides resolving the nature of the TAI phase, the present work intends to showcase the efficiency and accuracy of the calculations based on the non-commutative spin-Chern number.

\section{The model systems}

The computations will be carried out for models described by $H_0$ plus disorder. The generic form of the diagonal term is 
\begin{equation}
\epsilon({\bm k})=C-2D(2-\cos k_x -\cos k_y),
\end{equation}
and the minimal form of the matrix $\Gamma(k)$ was derived in Ref.~\onlinecite{Yamakage2010xr}:
\begin{equation}
\begin{array}{c}
\Gamma(k)=
i \Lambda \left (
\begin{array}{cc}
\sin k_x - i \sin k_y & 0 \\
0 & \sin k_x +i \sin k_y
\end{array}
\right ).
\end{array}
\end{equation}

The real space representation of $H_0$ can be constructed on a square lattice where each vertex ${\bm n}$ carries four quantum states $|{\bm n},\alpha,\sigma\rangle $, where $\alpha=\pm 1$ (= isospin) labels the $s$ or the $p$ character of the bands and $\sigma=\pm 1$ the spin up and down configurations. On the Hilbert space spanned by $|{\bm n},\alpha,\sigma\rangle$, the translational invariant Hamiltonian takes the form ($a$ = lattice spacing):
\begin{equation}
\begin{array}{c}
H_0=\frac{D}{a^2}(t_{1,0}+t_{-1,0}+t_{0,1}+t_{0,-1}-4+C/D) \medskip \\
+\frac{A}{2ia}\hat{\sigma}(t_{1,0}-t_{-1,0})r_\alpha+\frac{A}{2a^2}\hat{\alpha}(t_{0,1}-t_{0,-1})r_\alpha \medskip \\
+\frac{B}{a^2}\hat{\alpha}(t_{1,0}+t_{-1,0}+t_{0,1}+t_{0,-1}-4+M/B) \medskip \\
+\frac{\Lambda}{2a}\hat{\sigma}(t_{1,0}-t_{-1,0}) r_\sigma+\frac{\Lambda}{2ia}\hat{\alpha}(t_{0,1}-t_{0,-1}) r_\sigma,
\end{array}
\end{equation}
where $t_{m,k}$, $\hat{\sigma}$, $\hat{\alpha}$, $r_{\alpha}$ and $r_{\sigma}$ are the translations, spin, isospin and flipping operators defined below:
\begin{equation}
\begin{array}{c}
t_{m,k}|n_1,n_2,\alpha,\sigma\rangle = |n_1+m,n_2+k,\alpha,\sigma\rangle, \medskip \\
\hat{\sigma} |{\bm n},\alpha,\sigma\rangle = \sigma |{\bm n},\alpha,\sigma\rangle, \ \hat{\alpha}|{\bm n},\alpha,\sigma\rangle = \alpha |{\bm n},\alpha,\sigma\rangle, \medskip \\
r_\sigma |{\bm n},\alpha,\sigma\rangle = |{\bm n},\alpha,-\sigma\rangle, \ r_\alpha|{\bm n},\alpha,\sigma\rangle = |{\bm n},-\alpha,\sigma\rangle.
\end{array}
\end{equation}

When the parameters are given the values:\cite{Koenig:2008so} $A=364.5$ meV nm, $B=-686$ meV nm$^2$, $C=0$, $D=-512$ meV nm$^2$, $\Lambda=0$ and $a=5$ nm, the Hamiltonian $H_0$ accurately reproduces the relevant band structure of the properly tuned HgTe/CdTe wells in the QSH regime. Our calculations will be carried with these realistic parameter values but also with the theoretical values: $A=1$, $B=1$, $C=0$, $D=0$, $\Lambda=0$ (and $\Lambda=0.5$), and $a$ set to one. These later theoretical values have been used in Ref.~\onlinecite{Yamakage2010xr}, which contains one of the most accurate calculations to date for disordered topological insulators. For this reason, we decided to use Ref.~\onlinecite{Yamakage2010xr} for comparison and for assessing the accuracy of our calculations. 

The computations with disorder will be done with the Hamiltonian:
\begin{equation}\label{DisorderH}
H_\omega = H_0+ W \sum_{{\bm n},\alpha,\sigma} \omega_{{\bm n},\alpha} |{\bm n},\alpha,\sigma \rangle \langle {\bm n},\alpha,\sigma|,
\end{equation}
where $\omega_{{\bm n},\alpha}$ are random amplitudes uniformly distributed in the interval $[-\frac{1}{2},\frac{1}{2}]$. This is just a crude approximation of the disorder in QSH wells, where the disorder is probably mostly due to the random displacements of the atoms from the perfect crystalline structure. The leading physical effect of such displacements will be a change of the overlap integrals of the atomic orbitals. Consequently, the disorder in QSH wells will primarily occur in the hopping amplitudes. We are in the process of investigating such effects, but for now, we will follow the previous studies and work with the disordered Hamiltonian of Eq.~\ref{DisorderH}.

\section{The non-commutative spin-Chern invariant}

This invariant was discussed extensively in Refs.~\onlinecite{Prodan:2009oh,Prodan:2010cz,ProdanJPhysA2010xk} and here we will give only a brief account of it. Let $\hat{\sigma}_z$ be the operator $\hat{\sigma}_z|{\bm n},\alpha,\sigma\rangle$=$\sigma|{\bm n},\alpha,\sigma\rangle$ and suppose that an exact diagonalization was performed for $H_\omega$. Given a Fermi level, one can compute the projector $P_\omega$ onto the energy spectrum below $E_F$. Furthermore, one can diagonalize the operator $P_\omega \hat{\sigma}_z P_\omega$ and what he will find is an eigenvalue spectrum that is symmetric relative to the origin and contained in the interval $[-1,1]$. We denote by $P_\omega^\pm$ the spectral projector onto the positive/negative spectrum of $P_\omega \hat{\sigma}_z P_\omega$. Effectively, these projectors split the space of the occupied states into spin up and spin down sectors, and no $S_z$ conservation is required by the procedure. Now, the projectors $P_\omega^\pm$ fit into the non-commutative theory of the Chern invariant,\cite{BELLISSARD:1994xj} so one can define the non-commutative Chern numbers:
\begin{equation}\label{Chern1}
C_\pm=2\pi i \big \langle \mbox{tr}_0\big{ \{ }P_\omega^\pm \big{[} -i[\hat{x}_1,P_\omega^\pm],-i[\hat{x}_2,P_\omega^\pm] \big{]}\big{ \} } \big \rangle,
\end{equation}
where the outer angular parentheses signify disorder average, and tr$_0$ is the trace over the states at site ${\bm n}={\bm 0}$. Also, $\hat{{\bm x}}$ is the position operator. The non-commutative spin-Chern number is defined as 
\begin{equation}\label{spin-Chern}
C_s=\frac{1}{2}(C_+ - C_-).
\end{equation}

\begin{figure}
  \includegraphics[width=7cm]{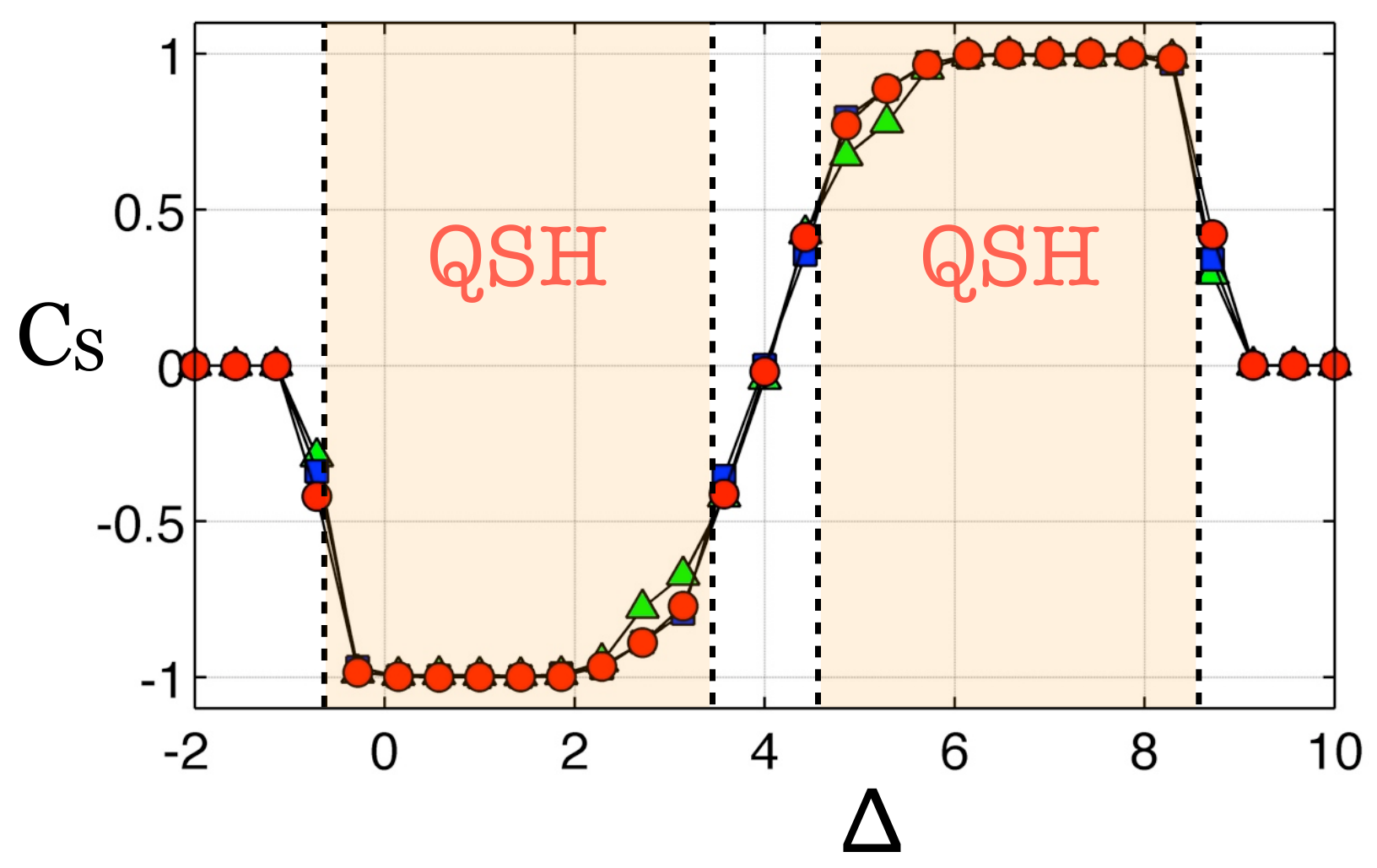}\\
  \caption{(Color online) The numerical values of the spin-Chern number $C_s$ as function of $\Delta$, when the other parameters were fixed at: $A=B=a=1$, $\Lambda=0$, $W=5.71$ and $E_F=0$. $C_s$ was computed for three lattices of increasing size: 30$\times$30 (triangles), 40$\times$40 (squares) and 50$\times$50 (circles). The shaded region indicates the region where $|C_s| \geq 0.5$, which is being identified with the QSH phase.}
 \label{SectionsSizeDepModel}
\end{figure}

An immediate consequence from Ref.~\onlinecite{BELLISSARD:1994xj} is that $C_s$ takes quantized values as long as the spectrum of $P_\omega \hat{\sigma}_z P_\omega$ is localized near the origin so that the matrix elements of $P_\omega^\pm$ decay sufficiently fast, more precisely, as long as: 
\begin{equation}
\sum_{\bm n} |{\bm n}|^2 \sum_{\alpha,\alpha',\sigma,\sigma'}  |\langle 0,\alpha,\sigma |P_\omega^\pm |{\bm n},\alpha',\sigma'\rangle|^2 <\infty.
\end{equation}
One can rigorously show that the region where $|C_s|=1$ is inside the  QSH phase, and that the region where $C_s=0$ is inside the normal insulating phase.\cite{ProdanJPhysA2010xk} Further investigations have shown that the delocalization of $P_\omega^\pm$ occurs simultaneously with the delocalization of the full projector $P_\omega$.\cite{ProdanJPhysA2010xk,Shulman2010cy} The reason for this is because the mobility gap of  $P_\omega \hat{\sigma}_z P_\omega$ is insensitive to spin-independent disorder, so its eigenstates near the origin remain localized as long as $P_\omega$ stays localized. The practical consequence of all this is that the region where $C_s$ takes the quantized values $\pm 1$ actually coincides with the QSH phase region and the region where $C_s$ takes the value 0  coincides with the normal insulating phase. The region where $C_s$ takes  non-quantized values can be identified with the metallic phase. These conclusions have been verified by extensive numerical computations.\cite{ProdanJPhysA2010xk,Shulman2010cy}

\begin{figure}
  \includegraphics[width=8.6cm]{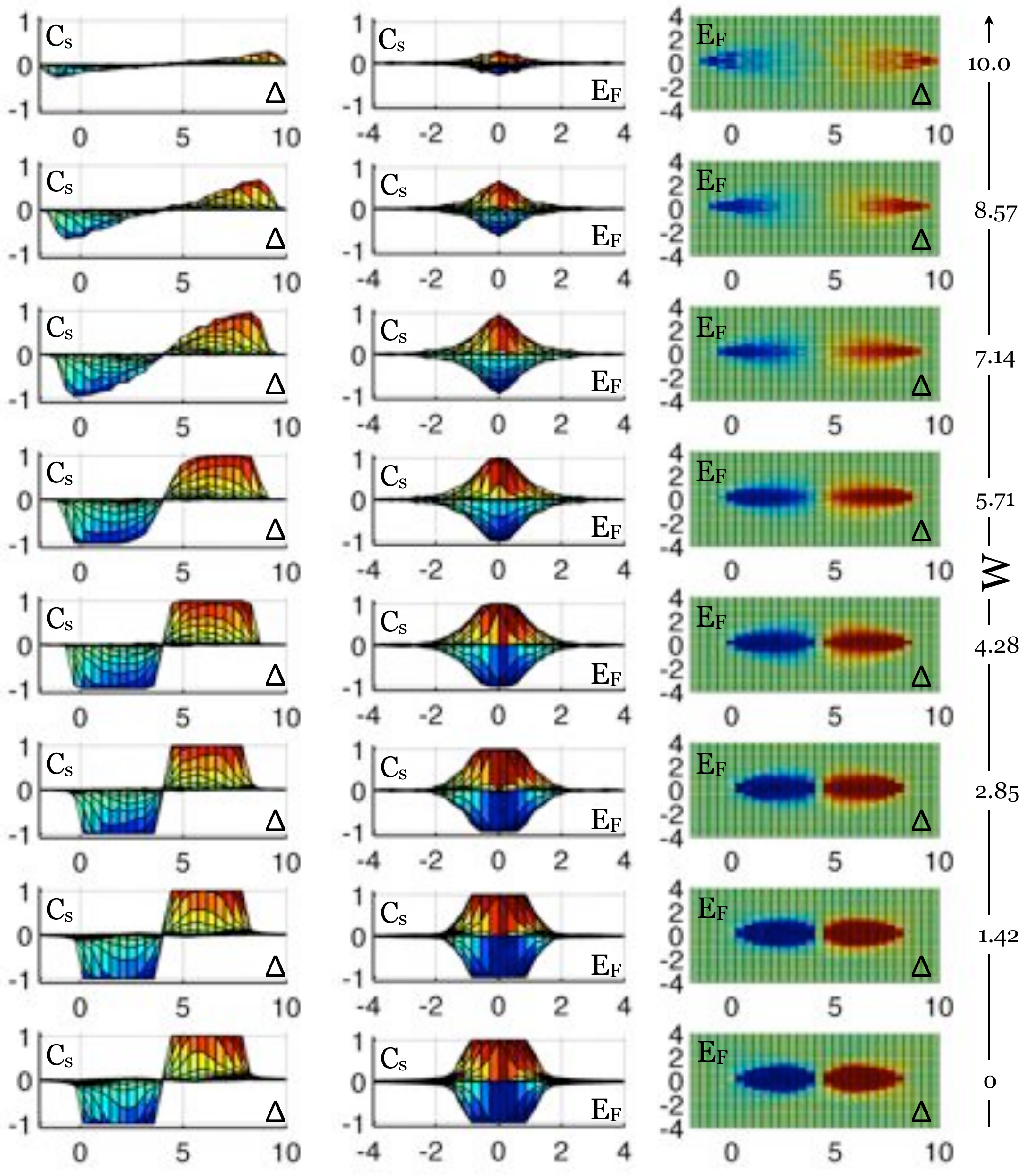}\\
  \caption{Global view of all our numerical data for $\Lambda=0$ (see Section~\ref{C1}), obtained on the 50$\times$50 lattice. $A$, $B$ and $a$ took the same values as in Fig.~\ref{SectionsSizeDepModel}. $C_s$ is represented by a surface plot, with $E_F$ and $\Delta$ on the horizontal axes and $C_s$ on the vertical axis. There is one surface plot for each $W$ value (shown on the right) we considered in our study. The surface plots are shown from three angles: sideway from $E_F$ direction (first column), sideway from $\Delta$ direction (second column) and from above (third column).}
 \label{50x50Model}
\end{figure}

The non-commutative formulas of Eq.~\ref{Chern1} makes sense only in the thermodynamic limit but in practice we can only work with finite size systems. For a finite  $N\times N$ square lattice with periodic boundary conditions, the following formula was derived:\cite{Prodan2010ew,ProdanJPhysA2010xk}
\begin{equation}\label{Chern2}
C_\omega^\pm=-\frac{2\pi i}{N^2} \mbox{Tr}\big \{P_\omega^\pm \big{[} -i\lfloor \hat{x}_1,P_\omega^\pm \rfloor,-i\lfloor \hat{x}_2,P_\omega^\pm \rfloor \big{]}\big \},
\end{equation}
where the trace is over all the states, and
\begin{equation}
\begin{array}{c}
\lfloor \hat{x}_i,P_\omega^\pm\rfloor =i \sum_{m=1}^Q c_m \times \medskip \\
(e^{-\frac{2\pi i}{N}m\hat{x}_i}P_\omega^\pm e^{\frac{2\pi i}{N}m\hat{x}_i}-e^{\frac{2\pi i}{N}m\hat{x}_i}P_\omega^\pm e^{-\frac{2\pi i}{N}m\hat{x}_i}).
\end{array}
\end{equation} 
Above, $Q$ is taken of the order of $N/2$ and the $c_m$'s are solutions of the following linear system of equations:
\begin{equation}
\begin{array}{c}
\hat{A}
\left (
\begin{array}{c}
c_1\\c_2\\ \ldots\\c_Q
\end{array}
\right)
=\frac{N}{\pi}
\left (
\begin{array}{c}
1 \\0\\ \ldots\\0
\end{array}
\right), \ \ A_{ij}=j^{2i-1}.
\end{array}
\end{equation}
With these choices, the finite size formulas from Eq.~\ref{Chern2} converges exponentially fast to the exact formulas given in Eq.~\ref{Chern1}, as long as the Fermi level is in a mobility gap. The exponential convergence slows down when the Fermi level nears a mobility edge. Consequently, the most challenging part of the calculations is to achieve convergence near the mobility edges, same as to say, near the phase separation lines.  

\begin{figure}
  \includegraphics[width=8.6cm]{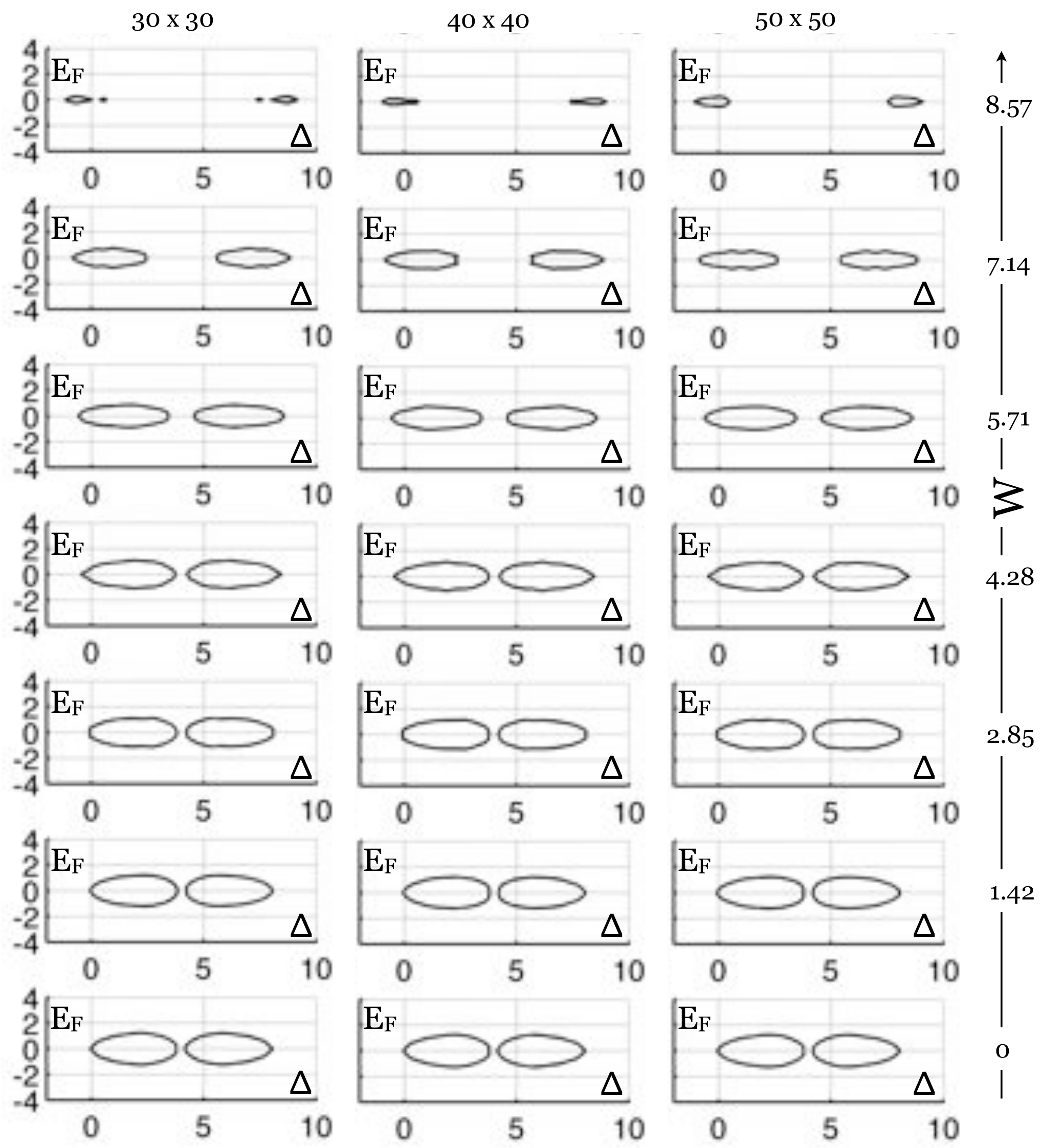}\\
  \caption{The contour levels $C_s=0.5$ obtained from the surface plots of Fig.~\ref{50x50Model} and from similar data obtained on the 30$\times$30 and 40$\times$40 lattices. When the calculations are converged with the size of the lattice, the region inside these contours contains the QSH phase.}
 \label{ModelContourSizeDep}
\end{figure}

From a numerical point of view, Eq.~\ref{Chern2} is straightforward to implement. Indeed, note that the matrix elements of $\lfloor \hat{x}_i,P_\omega^\pm\rfloor$ take the simple form:
\begin{equation}
\begin{array}{c}
\langle {\bm n},\alpha,\sigma|\lfloor \hat{x}_i,P_\omega^\pm\rfloor|{\bm n}',\alpha',\sigma'\rangle = \sum\limits_{m=1}^Q c_m \times \medskip \\
 \sin\left(\frac{2\pi i m(n_i-n'_i)}{N}\right) \langle {\bm n},\alpha,\sigma|P_\omega^\pm |{\bm n}',\alpha',\sigma'\rangle.
\end{array}
\end{equation}
Also note that $C_\omega^\pm$ are self-averaging, so the $\omega$-dependence becomes weaker and weaker as the volume is taken larger and larger. In practice, we still need to average over a few disorder configurations in order to obtain relatively smooth data. To summarize, the computation will proceed as follows:
\begin{itemize}
\item Chose a value for $\Delta$, $E_F$ and $W$. 
\item Generate a random potential and diagonalize the resulting $H_\omega$.
\item Construct the projector $P_\omega$ onto the states with energy below $E_F$.
\item Construct and diagonalize the operator $P_\omega \hat{\sigma}_z P_\omega$.
\item Construct the projectors $P_\omega^\pm$ onto the positive/negative spectrum of $P_\omega \hat{\sigma}_z P_\omega$.
\item Evaluate the finite size formulas Eq.~\ref{Chern2}.
\item Compute the spin-Chern number using Eq.~\ref{spin-Chern}.
\item Analyze the spin-Chern number data.
\end{itemize}

We end this section by mentioning that, for time-reversal symmetric models, the spin-Chern number modulo 2 coincides\cite{Prodan:2009oh,ProdanJPhysA2010xk} with the ${\bm Z}_2$ invariant introduced by Kane, Mele and Fu in Refs.~\onlinecite{Kane:2005zw} and \onlinecite{Fu:2006ka}. For example, the analytic calculations of Ref.~\onlinecite{LiPRB2010uv} showed explicitly that, for a certain $S_z$-nonconserving model in the clean limit, the kernel appearing in the computation of the spin-Chern number is identical to the Pfaffian function used in the computations of the $Z_2$ invariant (in the spirit of Ref.~\onlinecite{Kane:2005zw}) for the same model, performed in Refs.~\onlinecite{ShanNJP2010vy,LuPRB2010cu}. For this analytically solvable model, the $Z_2$ and the spin-Chern invariants were explicitly shown to give identical phase diagrams. 
 
\section{Model calculations and analysis}

In this section we set the parameters to the theoretical values mentioned above: $A=B=a=1$ and zero in rest, except for $\Lambda$ which will be given the values 0 and 0.5. When doing so, the model becomes particle-hole symmetric and the phase diagram becomes mirror symmetric relative to the plane $\Delta=4$. This is of great help when computing the 3D phase diagram, since we only need to cover the $\Delta \leq 4$ and $E_F\leq 0$ region (one quarter of the whole parameter space). This is what we will do and, once the calculations are completed, we reconstruct the full phase diagram by using these mentioned symmetries.

\begin{figure}
  \includegraphics[width=6cm]{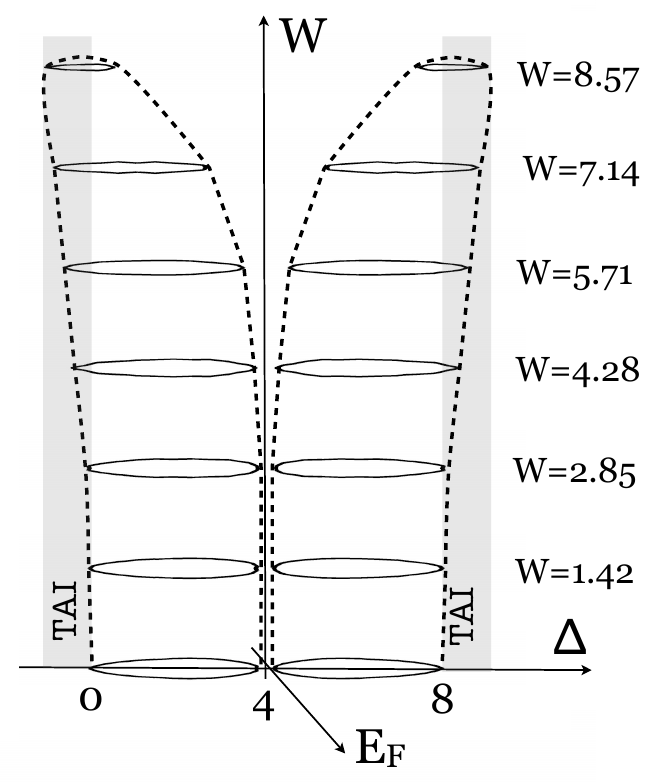}\\
  \caption{The 3D phase diagram of the model with $\Lambda=0$, as derived from the contours of Fig.~\ref{ModelContourSizeDep}. The 3D region delimitated by the contours and the dotted line contains the QSH phase. The shaded region indicates the region which contains the so called topological Anderson insulating phase. The dotted lines were obtain by extrapolation.}
 \label{ModelPhaseDiag}
\end{figure}

\subsection{$S_z$-conserving calculations}\label{C1}

We set here $\Lambda$ to zero, in which case $S_z$ commutes with the Hamiltonian and the model decouples into two Chern insulators, one on the spin-up sector and the other one on the spin-down sector. The model now belongs to the unitary class so one expects to see a sharp transition between the QSH and normal insulating phases.\cite{Onoda:2007xo,Prodan2010ew,ProdanJPhysA2010xk,Yamakage2010xr} The states strictly between the QSH and normal insulating phase are necessarily delocalized.\cite{BELLISSARD:1994xj,ProdanJPhysA2010xk}

\begin{figure}
  \includegraphics[width=8cm]{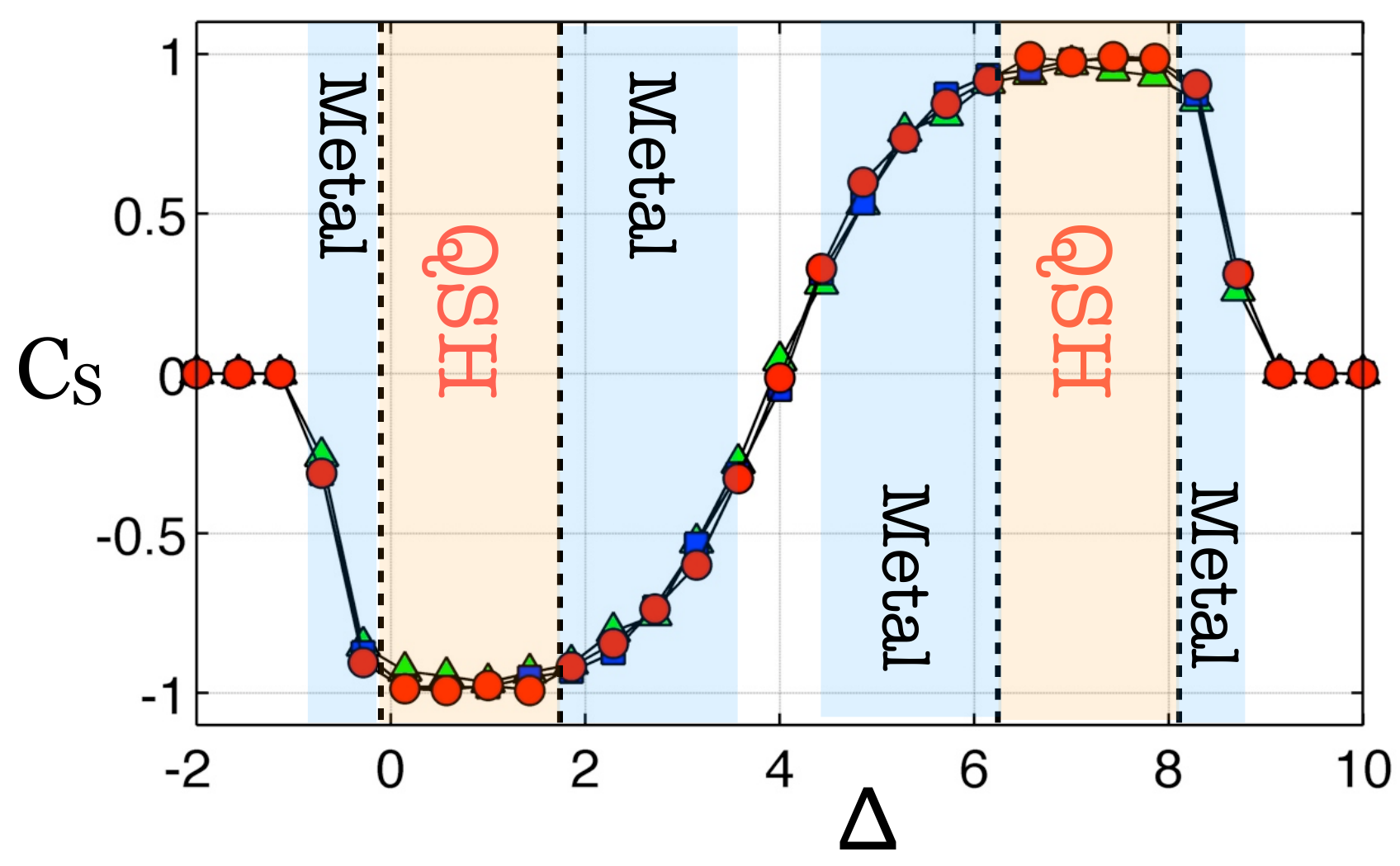}\\
  \caption{(Color online) The numerical values of the spin-Chern number $C_s$ as function of $\Delta$, when the other parameters were fixed at: $A=B=a=1$, $\Lambda=0.5$, $W=5.71$ and $E_F=0$. $C_s$ was computed for three lattices of increasing size: 30$\times$30 (triangles), 40$\times$40 (squares) and 50$\times$50 (circles). The various shaded regions indicate the regions where a) $1.5\leq |C_s| \leq 0.9$, which is being identified with the metallic phase and b) $|C_s| \geq 0.9$, which is being identified with the QSH phase.}
 \label{SectionsSizeDepModelRa}
\end{figure}

We first look in more detail at the numerical results. For this, we fix $W=5.71$ and $E_F=0$ and examine the values of $C_s$ as $\Delta$ is varied from $-2$ to 10.  Graphs of these values are shown in Fig.~\ref{SectionsSizeDepModel} for lattices of increasing sizes: 30$\times$30, 40$\times$40 and 50$\times$50. The level of disorder is high, the value $W=5.71$ being comparable with the width of the clean bands and much larger than the occurring insulating gaps. Consequently, the insulating gaps are completely filled with dense insulating spectrum in all calculations presented in Fig.~\ref{SectionsSizeDepModel}. $C_s$ was averaged over four disordered configurations. We have not considered larger sizes or additional disordered configurations because the resulting phase diagram is already well converged and the averaged $C_s$ appears smooth. As one can see from Fig.~\ref{SectionsSizeDepModel}, there are regions where $C_s$ takes quantized values of 0 and $\pm 1$, but also regions where $C_s$ takes non-quantized values. The regions with $C_s=0$ can be safely regarded as corresponding to the normal insulating phase, while those with $|C_s|= 1$ as corresponding to the QSH phase. It is known analytically that $|C_s|$=1 implies existence of edge states.\cite{Prodan:2009lo,Prodan:2009mi} To resolve the region where $C_s$ takes non-quantized values, which occurs precisely near the phase boundaries as discussed above, we relay on the behavior of $C_s$ when increasing the size of the lattice. By examining each point of the diagram, we see that, for specific values of $\Delta$, $C_s$ moves towards $\pm 1$, and, for other values of $\Delta$, $C_s$ moves towards 0. Thus, the size dependence of $C_s$ gives us a practical method to resolve the regions where the convergence is slow. 

\begin{figure}
  \includegraphics[width=8.6cm]{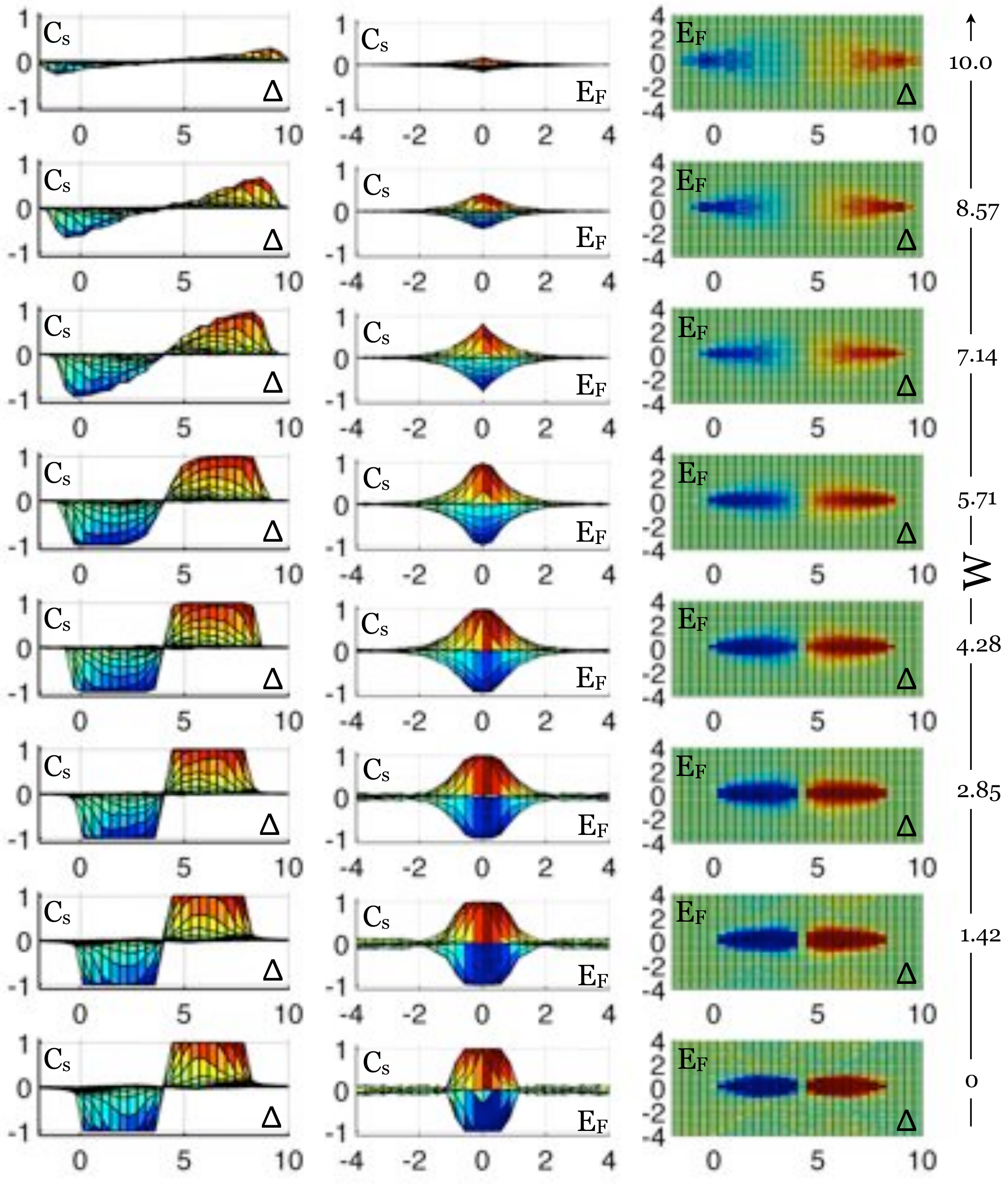}\\
  \caption{Global view of all our numerical data for $\Lambda=0.5$ (see Section~\ref{C2}), obtained on the 50$\times$50 lattice. $A$, $B$ and $a$ took the same values as in Fig.~\ref{SectionsSizeDepModelRa}. $C_s$ is represented by a surface plot, with $E_F$ and $\Delta$ on the horizontal axes. There is one surface plot for each $W$ value (shown on the right) we considered in our study. The surface plots are shown from three angles: sideway from $E_F$ direction (first column), sideway from $\Delta$ direction (second column) and from above (third column).}
 \label{50x50ModelRa}
\end{figure}

As an empirical fact, we found that the points above/below the values $\pm 0.5$ move up/down when the size of the lattice is increased, while the other points move down/up. Therefore, with a good measure, we can take the points where $|C_s|$=0.5 as the phase boundary between the QSH and normal insulating phases. This criterion is exemplified on the data shown in Fig.~\ref{SectionsSizeDepModel}, which then leads to the 1D phase diagram shown in the same figure.

Fig.~\ref{50x50Model} gives a global view of the data obtained on the 50$\times$50 lattice, by plotting $C_s$ as a function of both $\Delta$ and $E_F$ for several disorder strengths $W$. All our conclusions in this section are based on this data. From Fig.~\ref{50x50Model}, the reader can get a sense of how sharp are the regions where $C_s$ takes quantized values, how precise this quantization is (it is very precise),  and what is the extent of the regions where the calculations are not fully converged. From this global mapping we can extract the 3D phase diagram of the model. To do that, for each surface plot in Fig.~\ref{50x50Model}, we draw the contours corresponding to the level values $|C_s|$=0.5 and the results are shown in Fig.~\ref{ModelContourSizeDep}. We have placed the results for the different lattice sizes near each other so that one can examine the convergence of the phase diagram. It is quite evident that the contours are well converged with the size of the system for $W$ values up to 7.14. The computed QSH phase still has a small growth for $W$=8.57 as the size is increased. But overall, we believe the 50$\times$50 lattice calculations gives a fairly well converged 3D phase diagram of the model, which is shown in Fig.~\ref{ModelPhaseDiag}.

The phase diagram shown in Fig.~\ref{ModelPhaseDiag} is in excellent quantitative agreement with the data reported in Ref.~\onlinecite{Yamakage2010xr} (where only the section $E_F=0$ was examined). As already noted in Ref.~\onlinecite{Yamakage2010xr}, the phase boundary is strongly reshaped by disorder; the QSH phase is monotonically downsizing and drifts away from the $\Delta=4$ plane as $W$ is increased. Because of these particularities, the QSH phase extends in the regions $\Delta<0$ and $\Delta>8$, which correspond to the normal insulating phase when $W$=0. We dubbed this region the TAI region, because here is where the TAI phase will be observed if one would analyze only a slice (for example $\Delta$=-0.5) of the 3D phase diagram. But now it becames clear that the TAI phase is not a new phase but is part of the QSH phase, whose phase boundary was strongly reshaped by the disorder.

\begin{figure}
  \includegraphics[width=8.6cm]{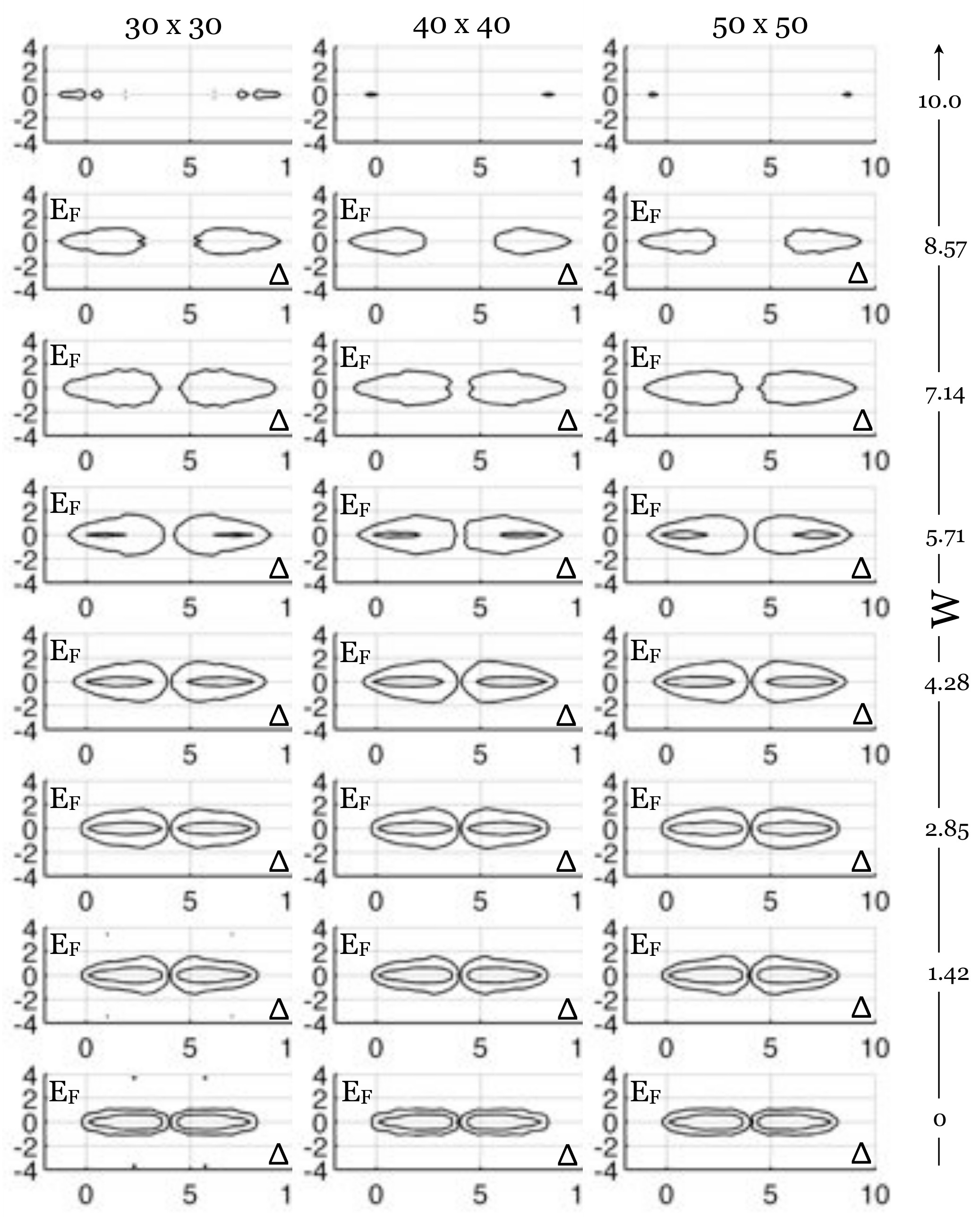}\\
  \caption{The contour levels $|C_s|=0.15$ and $|C_s|=0.9$ obtained from the surface plots of Fig.~\ref{50x50ModelRa} and from similar data obtained on the 30$\times$30 and 40$\times$40 lattices. The contours are easy to distinguish because the $0.9$ level contours are always inside the $0.15$ level contours. The regions inside the $0.9$ level contours give an approximate representation of the QSH phase, and the regions between the $0.15$ and $0.9$ level contours give an approximate representation of the metallic phase.}
 \label{ModelRaContourSizeDep}
\end{figure}

\subsection{$S_z$-nonconserving calculations}\label{C2}

Here we set the parameters at $A$=$B$=$a$=1 and $\Lambda$=0.5, so that a large $S_z$ non-conserving potential is present. In this case, the model belongs to the symplectic class, so one expects to see a metallic phase in between the QSH and normal insulating phases.\cite{ProdanJPhysA2010xk} The task of computing the phase diagram is much more difficult now, but nevertheless can be accomplished using the spin-Chern number.

\begin{figure}
  \includegraphics[width=5.5cm]{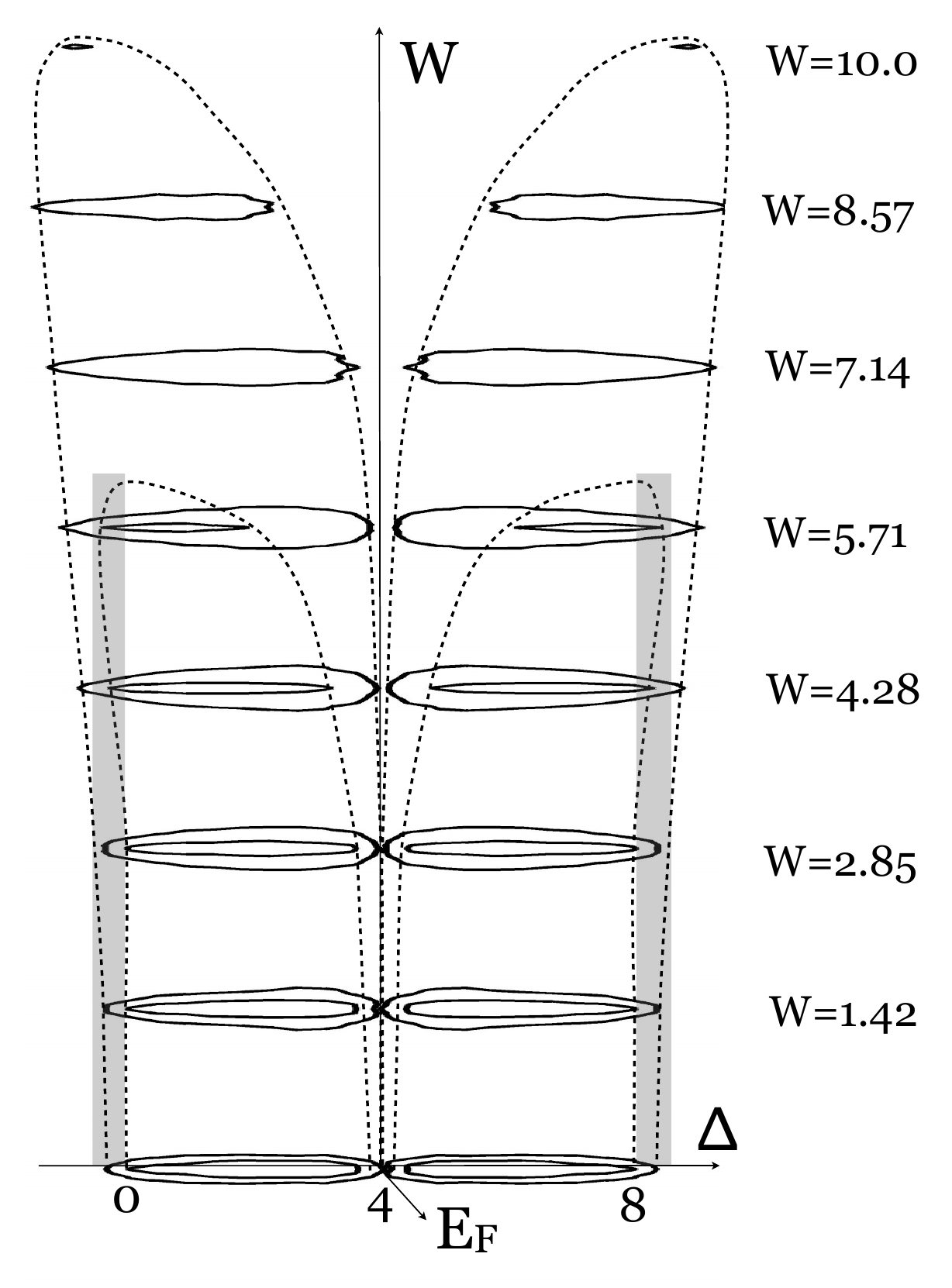}\\
  \caption{The 3D phase diagram of the model with $\Lambda=0.5$, as derived from the contours of Fig.~\ref{ModelRaContourSizeDep}. The 3D region delimitated by the inside contours and the dotted line contains the QSH phase. The 3D region between the contours and the dotted lines lines contains the metallic phase. The shaded region indicates the region which contains the so called topological Anderson insulating phase. The dotted lines were obtained by extrapolation.}
 \label{ModelRaPhaseDiag}
\end{figure}

Let us explore the numerical data in detail. As before, we fix two parameters: $W$=5.71 and $E_F$=0, and then examine the values of $C_s$ as $\Delta$ is varied from $-2$ to 10. The values are graphed in Fig.~\ref{SectionsSizeDepModelRa}, for three lattice sizes: 30$\times$30, 40$\times$40 and 50$\times$50. Again, one can see regions where $C_s$ takes quantized values $\pm 1$ and 0, which can be safely regarded as belonging to the QSH and normal insulating phases, respectively. There are also regions where $C_s$ doesn't take quantized values. The metallic phase is contained within this region, but of course there will be an inherent uncertainty in establishing the exact boundaries of the phases. 

Our criterion for phase delimitation will be as follows. We identify the QSH phase with the region where $|C_s|>0.9$ and the normal insulating phase with the region where $|C_s|<0.15$. The region where $0.15<|C_s|<0.9$ will be identified with metallic phase. If we apply this criterion to the data in Fig.~\ref{SectionsSizeDepModelRa}, we obtain the 1D phase diagram shown on top of the graphs in the same figure. The value $|C_s|$=0.15, chosen to define the boundary between the metallic and normal insulating phase, may appear high but this was the lowest value we could consider and still obtain smooth boundary separations. 

In Fig.~\ref{50x50ModelRa} we give a global view of the data obtained with a 50$\times$50 lattice. All our conclusions for this section are based on this data. It is quite obvious that the transition from the QSH phase to the normal insulating phase is much blurrier this time (compare with Fig.~\ref{50x50Model}), owing to the emergence of the metallic phase when $\Lambda \neq 0$. We now apply the above criterion and draw the contours corresponding to $|C_s|$=0.15 and 0.9 for each independent surface plot shown in Fig.~\ref{50x50ModelRa} (and similar plots generated with 30$\times$30 and 40$\times$40 lattices). The results are shown in Fig.~\ref{ModelRaContourSizeDep}. Here we again placed near each other the results for the three different lattice sizes: 30$\times$30, 40$\times$40 and 50$\times$50, so that one can judge the convergence of the phase boundaries with the size of the system. In our opinion, the contours in Fig.~\ref{ModelRaContourSizeDep} are well converged so we can proceed with the drawing of the phase diagram, which is shown in Fig.~\ref{ModelRaPhaseDiag}.  Our phase diagram is again in excellent quantitative agreement with the one reported in Ref.~\onlinecite{Yamakage2010xr}.  

Examining this 3D phase diagram, we observe the same trends seen in the previous calculations, with the QSH phase monotonically downsizing and moving away from the plane $\Delta=4$. The QSH phase region is smaller now when compared with the case $\Lambda=0$, but still there is a TAI region. In fact, as already mentioned in Ref.~\onlinecite{Yamakage2010xr}, there is a region of $\Delta$ values where, by just increasing the disorder in the system, one will observe a sequence of phase changes from normal insulator to metal, then to QSH insulator, then to metal and back to normal insulator.

\begin{figure}
  \includegraphics[width=8.6cm]{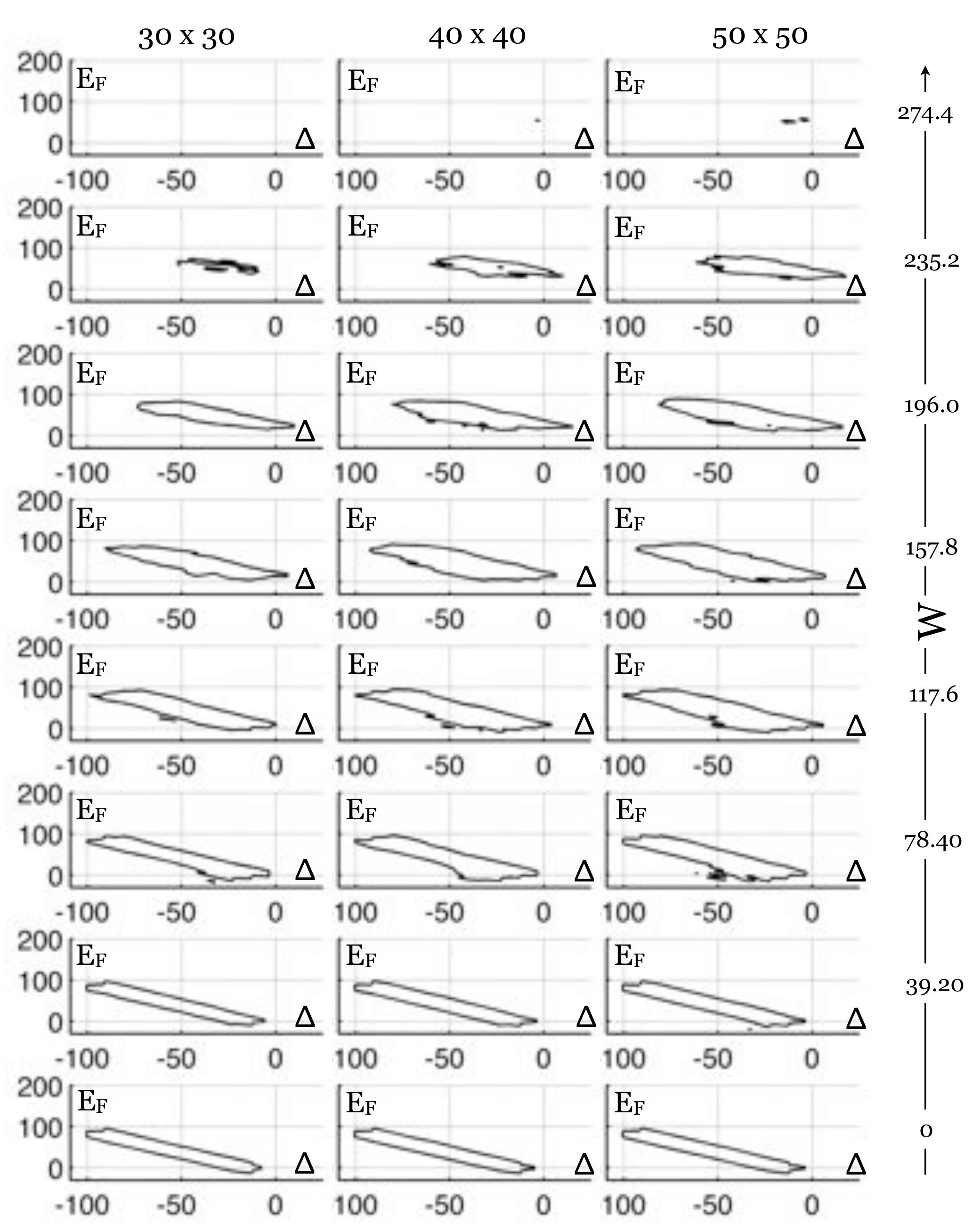}\\
  \caption{The contour levels $|C_s|=0.9$ obtained for the HgTe/CdTe QSH wlls computed on 30$\times$30 and 40$\times$40 lattices and for various disorder strenghts. When the calculations are converged with the size of the lattice, the region inside these contours contains the QSH phase.}
 \label{HgTeContourSizeDep}
\end{figure}

\section{Quantum wells in the spin-Hall regime}

Here we set the parameters so that the band structure of $H_0$ matches that of the HgTe/CdTe quantum well in the spin-Hall regime:\cite{Koenig:2008so}  $A$=364.5 meV nm, $B$=-686 meV nm$^2$, $C$=0, $D$=-512 meV nm$^2$, $\Lambda$=0 and $a$=5 nm. These same parameters were used in the theoretical studies of Refs.~\onlinecite{LiPRL2009xi,GrothPRL2009xi}, which we will compare with. Note that the $S_z$-nonconserving potential is being turned off in these calculations.

It was shown in these two mentioned references that the QSH phase region (and the one referred to as TAI) moves up in energy with the increase of the disorder amplitude. No such behavior was observed in our previous calculations, so we must conclude that this behavior is triggered by the presence of the diagonal term $\epsilon({\bm k})$. The self-consistent Born approximation analysis reported in Ref.~\onlinecite{GrothPRL2009xi} does not capture this fact; for example, Eq.~5 from this reference, giving the renormalized mass and chemical potential, implies that the lower edge of the mobility gap moves down when increasing $W$. Another aspect revealed by Refs.~\onlinecite{LiPRL2009xi,GrothPRL2009xi} is that the TAI region moves deeper and deeper into the normal insulating region of the clean limit when $W$ is increased. We will pay particular attention to these two aspects. 

\begin{figure}
  \includegraphics[width=8.6cm]{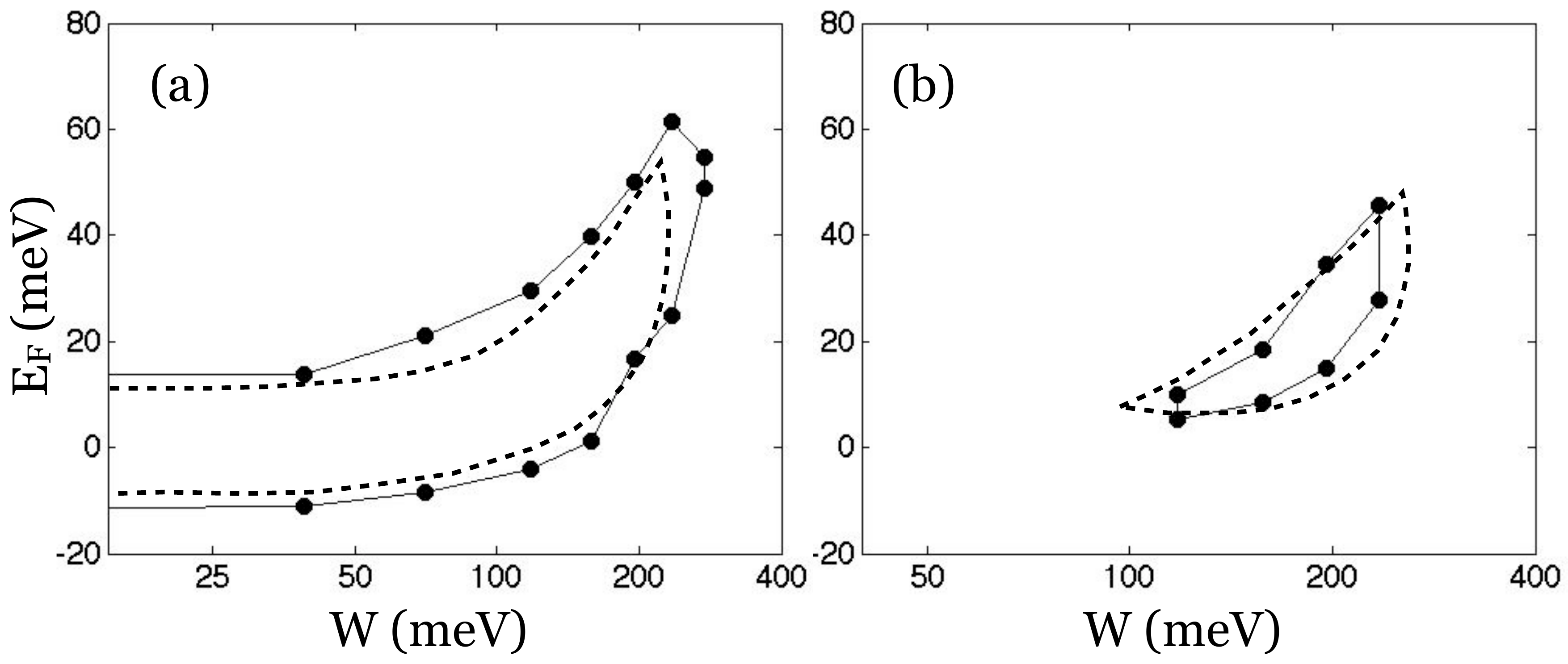}\\
  \caption{The solid dots and solid lines represent 2D slices of the 3D phase diagram shown in Fig.~\ref{SectionsSizeDepModelRa} (corresponding to the 50x50 lattice), taken at: (a) $\Delta=-13.5$ meV and (b) $\Delta=5.75$ meV. The dotted contour in (a) represents the boundary of the QSH phase, which we traced out from Fig.~2c of Ref.~\onlinecite{LiPRL2009xi}. The dotted contour in (b) represents the boundary of the TAI phase, which we traced out from Fig.~2f of Ref.~\onlinecite{LiPRL2009xi}.}
 \label{HgTeComp}
\end{figure}

We have repeated the calculation and analysis of the spin-Chern number along the steps detailed in the previous sections. In Fig.~\ref{HgTeContourSizeDep}, we show the phase separation lines between the normal and the QSH phase. The criterion adopted here is that QSH phase corresponds to the regions where $|C_s|>0.9$. Note that this is different from the criterion used in our model calculation, and there is a good reason for this. Because the QSH phase is being pushed at higher energies, the localization lengths of the localized states are much larger in the present calculations than those observed in the model calculations. As a result, the transition between the QSH and normal insulating phase is not as sharp as in the model calculations. By choosing the high value of $0.9$, we tried to make sure that the computed QSH phase is actually inside the exact QSH phase. In this way, we can be sure that, when the approximately computed QSH phase spills out into the normal phase, the exact QSH phase does the same thing.

Fig.~\ref{HgTeContourSizeDep} reveals that no disconnected phase regions are emerging when increasing $W$. Instead, the boundary between the QSH and normal phases moves continuously deeper and deeper inside the $\Delta>0$ region. In the same time, the boundary moves up in energy. This behavior is very similar to the behavior of TAI phase mentioned above. Refs.~\onlinecite{LiPRL2009xi,GrothPRL2009xi} drew phase diagrams in the $(W,E_F)$ plane corresponding to $\Delta=-10$ meV and $\Delta=1$ meV. These diagrams can now be viewed as slices of the 3D phase diagram shown in Fig.~\ref{HgTeContourSizeDep}, taken at the appropriate planes  $\Delta=-10$ and 1 meV. We want to show explicitly that this is the case. Because of the way we discretized $\Delta$ in our numerical calculations, we cannot consider these same $\Delta$ values, and instead we will draw the sections corresponding to $\Delta=-13.5$ meV and $\Delta=5.75$ meV. These sections are shown in Fig.~\ref{HgTeComp}, together with the QSH and TAI phase boundaries from Refs.~\onlinecite{LiPRL2009xi,GrothPRL2009xi}. 
The quantitative agreement seen in Fig.~\ref{HgTeComp} leaves no doubt that the phase diagrams reported in Refs.~\onlinecite{LiPRL2009xi,GrothPRL2009xi} are just particular slices of the 3D phase diagram reported in Fig.~\ref{HgTeContourSizeDep}.

Nevertheless, the results show that a material can become a QSH insulator by just increasing the disorder of the crystalline structure. It is a well established fact that a trivial Anderson insulator will eventually emerge at very large disorder strengths\cite{Aizenmann1993uf} so, at least in this regime, one can expect a monotonic shrinkage of the QSH part of the phase diagram as the disorder is increased. But at lower disorder strengths, we now know that the phase diagram can display complex trends. For the particular model of the HgTe wells and the particular disorder considered in our study, we actually see an initial expansion of the QSH phase, accompanied by a drift. For weak disorder strengths, one can use a perturbative calculation to analyze the trends, and now it is well known from Ref.~\onlinecite{GrothPRL2009xi} that the first correction to the Dirac mass has a positive sign, hence a drift to right of the QSH phase is expected. What was also interesting in Ref.~\onlinecite{GrothPRL2009xi} was that the perturbative analysis remained accurate up to quite large disorder strengths, so in some sense the general trends are set by the behavior at very weak disorder. It remains, however, to be explored how universal these conclusions are. For example it will be interesting to see what happens when a different kind of disorder is considered, like the one affecting the hopping terms as discussed in one of our sections.

\section{Conclusions}

We have computed the phase diagram of a Quantum spin-Hall model Hamiltonian in the 3D parameter space of Dirac mass, Fermi energy and disorder strength. The analysis was based on the computation of the spin-Chern number, which was shown to display quantized values in the presence of large disorder and large $S_z$-nonconserving interactions. Working with certain theoretical values, we showed first that our computations are in excellent quantitative agreement with the 2D phase diagrams reported in Ref.~\onlinecite{Yamakage2010xr}. The 3D phase diagrams confirm the absence of disconnected phase pockets and instead reveal a strong disorder-induced deformation of the phase boundaries. Secondly, working with parameters appropriate for a HgTe/CdTe quantum well tuned for the quantum spin-Hall effect, we observed a similar disorder-induced deformation of the phase boundaries and no disconnected phase pockets. We have shown that the 2D phase diagrams reported in Refs.~\onlinecite{LiPRL2009xi,GrothPRL2009xi} correspond to slices appropriately taken from our 3D phase diagram. This shows explicitly that the so called TAI phase is not a new and distinct phase. Instead, TAI is part of the QSH phase whose boundary is strongly deformed as the disorder amplitude is increased. 

\begin{acknowledgments} 
This research was supported by a Cottrell award from the Research Corporation for Science Advancement and by the office of the Provost of Yeshiva University. 
\end{acknowledgments}

%\bibliography{../../../TopologicalInsulators}

%Merlin.mbs v4.21 2009-07-09.
%

\end{document}